\documentclass[a4paper, 11 pt, onecolumn, draftcls]{ieeeconf}   
\IEEEoverridecommandlockouts                              

\usepackage{etoolbox}
\usepackage{float}
\usepackage{xcolor}
\usepackage{multirow}
\makeatletter
\def\@ieeeconfsetup{%
  \IEEEtriggeratref{0}%
  \IEEEtriggercmd{}%
  \normalsize\normalfont\rmfamily
  \setlength{\columnsep}{0.2in}%
}
\makeatother

\usepackage[pagewise, switch]{lineno}

\usepackage{graphicx}
\usepackage{mathtools}
\usepackage{mathptmx} 
\usepackage{mathptmx} 
\usepackage{amsmath} 
\usepackage{amssymb}  
\usepackage{hyperref}
\usepackage[acronym]{glossaries} 

\newcommand{\comp}[1]{\textbf{\texttt{#1}}}

\title{\LARGE \bf
Convolutional-Neural-Networks for Deanonymisation of\\ I2P Traffic}

\author{Luca Rohrer$^{1}$, Konrad Bächler$^{2}$, and Dieter Arnold$^{1}$
\thanks{$^{1}$L. Rohrer and D. Arnold are with the Lucerne University of Applied Sciences and Arts, Department of Computer Science, CH-6434 Rotkreuz, Switzerland.
        {\tt\small rohrer.luca@gmail.com} and  {\tt\small dieter.arnold@hslu.ch}} %
\thanks{$^{2}$K. Bächler is with diva.exchange, Switzerland.
        {\tt\small konrad@diva.exchange}}%
}

\begin{document}

\newacronym{ml}{ML}{Machine Learning}
\newacronym{cnn}{CNN}{Convolutional Neural Network} 
\newacronym{cnns}{CNNs}{Convolutional Neural Networks} 
\newacronym{rnn}{RNN}{Recurrent Neural Networks} 
\newacronym{i2p}{I2P}{Invisible Internet Project} 
\maketitle
\thispagestyle{plain}
\pagestyle{plain}

\begin{abstract}
This study investigates the potential for deanonymizing services within the \gls{i2p} network through passive traffic analysis and machine learning techniques. The primary objective is to identify distinctive patterns in \gls{i2p} traffic despite the encryption of its payload. To achieve this, a controlled laboratory environment was established to generate synthetic \gls{i2p} traffic, providing a training dataset for machine learning models. Furthermore, Fano's inequality is employed to perform a theoretical analysis of anonymous data transmission in mix networks such as \gls{i2p}, thereby supporting a data-driven approach to uncover causal relationships. In computer experiments, advanced deep learning methods - particularly \gls{cnns} - are applied within the laboratory \gls{i2p} network, and their effectiveness is further evaluated using real-world traffic data. The results indicate that the proposed methodologies do not compromise the anonymity guarantees of the \gls{i2p} network.
\end{abstract}


\section{Introduction}

The Invisible Internet Protocol (I2P) is a privacy-focused, fully distributed overlay network designed to provide communication anonymity over the internet. It achieves this by routing traffic through multiple peers using end-to-end encryption and an enhanced form of onion routing. \gls{i2p} functions as a transport-level anonymization layer, concealing both origin and destination of communications from passive and active observers \cite{i2pwy}. 

Anonymity in networks signifies the lack of identifiable links. Since absence cannot be directly observed, it cannot be definitively verified through deduction. Instead, various methods attempt to uncover links; if none succeed, communication is considered anonymous. Each failed attempt merely shows that anonymity has not yet been disproven. Following Popper's falsification principle \cite{popper}, positive proof of anonymity is impossible—its scientific validity relies on the persistent failure of attempts to refute it. Thus, claiming a network is anonymous is a provisional hypothesis, valid only as long as rigorous tests persistently fail to expose identifying connections.

Egger et al. \cite{egger} analyzed attacks against low-latency anonymity networks, including Tor and I2P. These attacks, also termed Sybil and Eclipse attacks, assume control of a majority of the network nodes. Since then, \gls{i2p} has been hardened against many such types of attack and a successful deanonymization of \gls{i2p} with limited control of the network nodes is to the best of our knowledge not known as of today. Known vulnerabilities of the protocol are listed in the threat model of \gls{i2p} \cite{tm}. Over time, the \gls{i2p} network has evolved into a mature anonymity system, yet its documentation remains fragmented, inconsistent, and often difficult to navigate—especially for researchers and practitioners seeking a clear theoretical understanding of how anonymity is achieved. 

The advancement of artificial intelligence (AI) and \gls{ml} has introduced new opportunities for passive attacks in anonymity networks. Shahbar and Zincir-Heywood successfully distinguished among websites, IRC chats, and \gls{i2p} Snark torrent services through traffic analysis and decision tree classification by recording and analyzing all device traffic \cite{shahbar}. Additionally, Montieri et al. demonstrated in \cite{montieri} that supervised \gls{ml} techniques can effectively classify data packets according to their respective anonymization networks, such as \gls{i2p}, Tor, and JonDonym.  

\gls{cnns} have become an effective tool for classifying network traffic and identifying protocols, even in scenarios where portions of the data are encrypted. Research indicates that \gls{cnns} can successfully detect patterns within packet structures, timing, and flow behaviors—features that are accessible independently of payload content. Several studies have investigated the application of \gls{cnns} in network traffic analysis. In \cite{lotfollahi2019deeppacket}, the authors integrate \gls{cnns} with autoencoders to classify encrypted traffic, enabling detailed differentiation between applications such as BitTorrent and Skype, as well as broader categories like FTP or peer-to-peer traffic. The combination of \gls{cnns} for spatial feature extraction with Long Short-Term Memory networks for modeling temporal dependencies is a recent promising approach for network protocol classification \cite{zhao2025cnn}. 

Our contribution is threefold: (1) we provide a concise and structured explanation of how \gls{i2p} achieves anonymity, grounded in the foundational concepts introduced by David Chaum’s mix networks; (2) we offer a theoretical framework that justifies the need for modern machine learning techniques for  analyzing anonymized traffic; and (3) we empirically evaluate the performance of \gls{cnns} and competing methods on real-world \gls{i2p} traffic data in order to assess their effectiveness in potential deanonymization.

\section{Anonymous Networks}
David Chaum’s 1981 paper \cite{chaum} is the intellectual foundation for anonymous networks and systems like \gls{i2p} and Tor. His work defined core concepts like mixing, pseudonymous communication, and untraceability — all of which are essential to both sender and receiver anonymity in modern anonymous overlay networks. 

\subsection{The Concept of Mix-nets}

The original concept of a mix-net based on Chaum is defined as an ordered chain of at least three network participants, referred to below as nodes, with the purpose of sending a message from the first node, the sender, to the last node, the recipient. The task of the nodes between the sender and the recipient is to forward the message.

Let us now define anonymity in simplified terms as: the forwarding nodes do not know the address (like an IP address) of either the sender or the recipient. To create a network which is in line with this definition of anonymity, each node in the network needs a key pair, a Public (Pub) and a Private (Priv) Key. Such a pair gets generated locally on each node using a well-known public-key cryptographic system \cite{chaum}.

Assume that each node stores this public key, together with the associated address, in a database which is fully replicated across the network. We refer to this database as the well-known network database. For the sake of clarity, we will ignore numerous issues at this point, such as the fact that writing to the network database must not breach anonymity or also that a full replication of the network database leads to significant network traffic. In such a system, each node knows all public keys of all nodes in the network and the corresponding address.

\begin{figure}[H] 
    \centering 
    \includegraphics[width=0.8\textwidth]{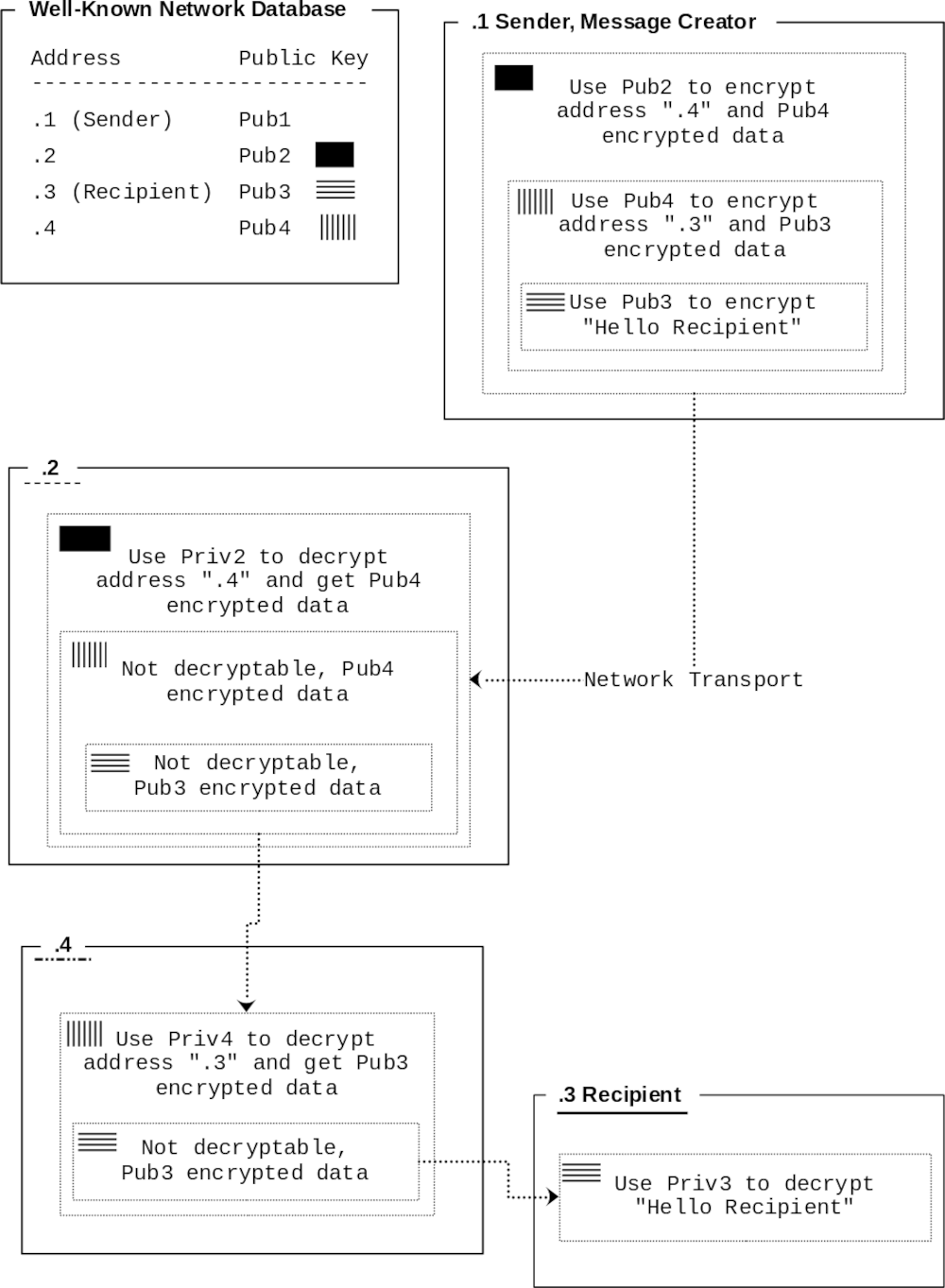} 
    \caption{Schema of a naive mix-net and the use of onion routing}
    \label{fig:onionexample} 
\end{figure}

Figure \ref{fig:onionexample} shows the conceptual functioning of a naive mix-net that provides a basic level of anonymity by using layered messages that are transported through a network. This concept is generally termed onion routing \cite{onionrouting}.

An important principle of onion routing is the use of all well-known public keys of nodes along a route defined by the sender. The public key of a node along a route is used to encrypt the address of the next node along the route. Applying this technique leads to a chain of nodes where every intermediate node along the route only knows the preceding and the following node.

Figure \ref{fig:onionexample} illustrates an example of how a core message, "Hello Recipient", is transmitted from Sender ".1" to Recipient ".3" within a mix-net.

The example, as shown in Figure \ref{fig:onionexample}, highlights several shortcomings:
\begin{itemize}
\item the recipient's address must be known by the sender;
\item the structure of the message allows conclusions about the length of the chain;
\item if a reply is to be possible, the recipient must be given the sender's address.
\end{itemize}

For these reasons, such a naive mix-net does not offer a useful degree of anonymity. That is why mix-net and onion routing have been further developed: among others, by \gls{i2p}, which has implemented the concept of unidirectional tunnels and aims to ensure a particularly high degree of anonymity.

\subsection{The Concept of Unidirectional Tunnels}

The concept of unidirectional tunnels arose from the further development of the mix-net and onion routing concepts. The concept of unidirectional tunnels is intended to solve the following key problems:
\begin{itemize}
\item all participants in a tunnel, including the sender and the recipient, remain anonymous,
\item the writing to the network database must not jeopardize the anonymity of nodes,
\item timing or fingerprinting attacks to deanonymize nodes without controlling the majority of the network are not successful.
\end{itemize}

The fundamental requirement for this form of anonymous communication is that each node generates two independent key pairs using a well-known public-key cryptographic system. It is crucial that these keys are written to the network database anonymously and in a completely uncorrelated manner. The well-known network database will then contain the following, uncorrelated, information:
\begin{itemize}
\item the first public key with the associated address of the network participant,
\item and the second public key together with a pseudonym, whereas the pseudonym might be a derivative of this second public key \footnote{I2P has decided to implement pseudonyms as so called "b32-addresses" which are derivatives of the public key}.
\end{itemize}

The first public key is solely used for onion routing instructions, whereas the second public key is used exclusively for end-to-end encryption of core messages.

An unidirectional tunnel is a chain of nodes and has an unchangeable direction of data flow. The sender uses outbound tunnels and the receiver uses inbound tunnels. Outbound tunnels protect the identity of the sender, and inbound tunnels protect the identity of the recipient.

Hence, outbound tunnels are defined and created by senders. To create such an outbound tunnel, the sender forms a stack of $1$ to $n$ addresses (such as IP addresses) randomly\footnote{Note: randomness might not be a desired property. \gls{i2p} implemented a specific peer selection process  \url{https://geti2p.net/en/docs/how/peer-selection}}  selected from the network database. This stack defines the exact route through the network. This is equivalent to the concept of onion routing shown in Figure \ref{fig:onionexample}. 

Inbound tunnels are defined and created by the recipients. The creation process is the same as for outbound tunnels. These inbound tunnels, together with the address of the inbound entry node and the pseudonym related to the second public key used for end-to-end encryption, are anonymously written to the network database via an outbound tunnel.

\begin{figure}[H] 
    \centering 
    \includegraphics[width=0.73\textwidth]{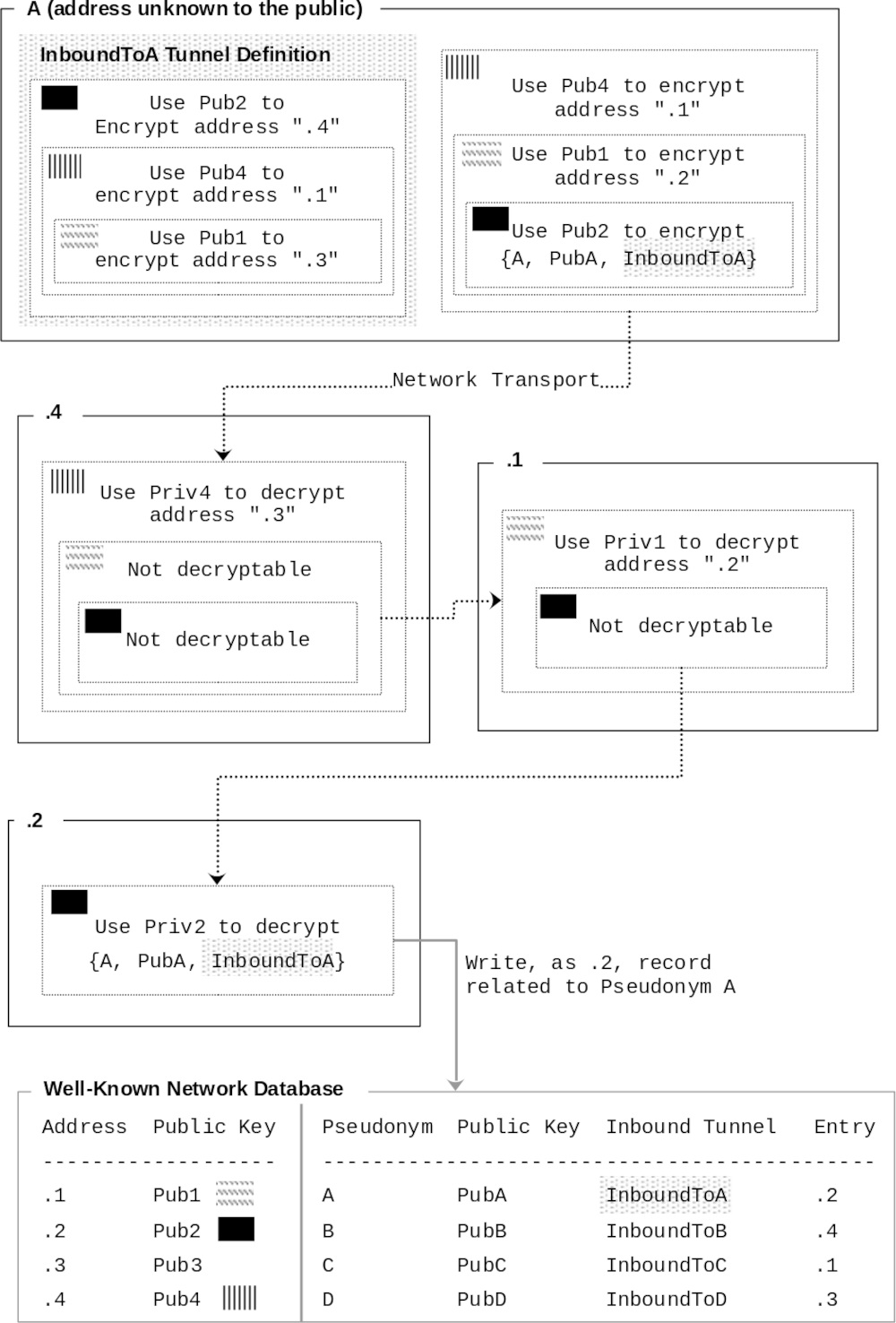} 
    \caption{Schema of Inbound Tunnel creation and the process of anonymous writing to the network database using an Outbound Tunnel}
    \label{fig:inboundexample} 
\end{figure}

Figure \ref{fig:inboundexample} extends the concept shown in Figure \ref{fig:onionexample} to illustrate how inbound tunnels are created and anonymously written to the network database using an outbound tunnel.

As shown in Figure \ref{fig:inboundexample}, no observer, apart from A, can create a logical relation between the address ".3" and the pseudonym "A" or between "Pub3", used for routing purposes, and "PubA" used for end-to-end encryption.

Inbound and outbound tunnels do not have to be the same length. The length and composition of unidirectional tunnels should also vary over time. Longer tunnels generally increase the degree of anonymity and cost transport time and energy. I2P, as an example, exposed these parameters to users, enabling them to configure the length and variability of tunnels.

An implementation must also provide answers to questions like:
\begin{itemize}
    \item What is the process to build persistent and re-usable tunnels?
    \item Which nodes are selected to be member of a tunnel?
    \item What is the lifetime of a tunnel?
    \item How does the network handle the evolution of cryptography over time?
\end{itemize}

Even if the implementations of anonymous networks differ: if a concept of an anonymous network is correctly implemented using unidirectional tunnels, the result is necessarily a completely trustless network. In a completely trustless network, there are additional challenges to solve, such as the need for the recipient of a message to be able to independently verify that the message has not been altered during transport, that the message has taken exactly the route intended by the sender and that the network is protected from resource abuse.

\subsection{The Invisible Internet Protocol (I2P)}

Based on the concept of unidirectional tunnels, an initial version of The Invisible Internet Protocol (I2P) was developed in the early 2000s \cite{i2np}. Over the following 20 years, the open source community continuously developed and tested \gls{i2p}. The first public commit to the Java version of I2P was on April 8, 2004\footnote{ \url{https://github.com/i2p/i2p.i2p/commit/77bd69c5e55428ec4c2db1589a6af1da15c10d05}}. The first public commit to the C++ version of \gls{i2p} (I2pd) came later on September 1st, 2013\footnote{ \url{https://github.com/PurpleI2P/i2pd/commit/bab26a5a14ee7a13014700ced66022a2769c385d}}. 

Like Tor, \gls{i2p} operates with low latency, which makes it susceptible to timing analysis and traffic correlation if an attacker controls substantial portions of the network. However, \gls{i2p} offers certain advantages over Tor, such as unidirectional tunnels.

Compared to Tor, I2P's smaller user base results in a reduced anonymity set, particularly in niche or smaller communities. Applications running on \gls{i2p} (including web servers, torrent clients, and messaging platforms) may still be vulnerable to data leaks. It is important to note that \gls{i2p} primarily secures the network layer and does not address vulnerabilities within application layers.

\section{Anonymity, Information Leakage, and Trustlessness}

Differential privacy \cite{dwork2006calibrating} is the gold standard for statistical data release, but is not well-suited for network anonymity. Its core mechanism of adding noise to query outputs is designed to protect individuals in a dataset, not to hide the endpoints of a live communication channel. Applying its principles directly to network packets would irreparably damage latency and throughput, rendering real-time communication impractical \cite{dpcom}.

Similarly, the concept of $k$-anonymity \cite{pfitzmann}, designed for released structured data (like a database table) is ill-suited in a dynamic network setting. The goal is to modify a dataset so that any individual in it cannot be distinguished from at least $k-1$ other individuals based on a set of "quasi-identifiers" (e.g., ZIP code, age, gender). Sophisticated traffic analysis attacks may compromise the privacy guarantees of $k$-anonymity in network flows. It is conceivable that a passive adversary can leverage timing, packet sequences, and background knowledge to isolate individual flows, violating anonymity despite the group size \cite{touchfromdistance}. 

\subsection{Information Leakage reduces Anonymity}
We denote by $\Theta$ the anonymity set, the set of the recipients among which our target node is hiding. The smaller this set is, the less anonymous our target node is. Without observing any traffic, the size of the anonymity set $\Theta$  equals $N-1$ where $N$ is the number of all nodes in the network. The eavesdropper can now collect information over time and potentially reduce that set. For a qualitative understanding of this process we resort to Fano's inequality \cite{fanoie}.

Fano's inequality is a fundamental result in information theory that provides a lower bound on the probability of incorrectly estimating a random variable in terms of mutual information between that random variable and its estimate. 

Let $X$ be a random variable taking values in a finite set of size $N$, let $Y$ be any observation, and $\hat{X}$ an estimate of $X$ (possibly dependent on $Y$). Then, we define the probability $P_e$ of not correctly estimating $X$ from observing $Y$ as $P_e := Pr[\hat{X}(Y)\neq X]$. Fano's inequality \cite{fanoie} amounts then to
\begin{equation}
     H(X\vert Y)\leq h(P_e) + P_e\cdot \log (N - 1)
    \label{eq:fano}
\end{equation}
where $H(X\vert Y)$ is the conditional entropy of $X$ given $Y$, and  $h(P_e)$ is the binary entropy function. Since $h(P_e)\leq 1$ and $N-1$ is the size of the anonymity set $\Theta$, we can rearrange the Fano inequality as
\begin{equation}
    P_e \geq \frac{H(X) - I(X;Y) - 1}{\log \vert \Theta \vert}.
    \label{eq:fanomi}
\end{equation}
where we made use of $H(X\vert Y)=H(X)-I(X;Y)$. 

The Fano inequality is used to prove impossibility results, i.e., no model, no estimator $\hat{X}(Y)$ can perform better than the expression on the right hand side of (\ref{eq:fanomi}). That expression is determined by the size of the anonymity set $\theta$, the uncertainty $H(X)$ about the recipient $X$ and the information (leakage) $I(X;Y)$ between $X$ and $Y$. A non-zero information leakage leads to a reduced  $P_e$.

If there is no communication at all, $Y$ leaks no information about $X$, then $I(X;Y)$ equals zero. So $P_e$ is bounded from below by the uncertainty about the recipient and the size of the anonymity set $\Theta$, i.e.
\begin{equation}
    P_e \geq \frac{H(X) - 1}{\log \vert \Theta \vert}.
\end{equation}
If all $N$ nodes are equally likely $H(X)=\log N$ and $P_e$ leads to $ P_e \geq \frac{\log N - 1}{\log (N -1)}$ and for $N$ sufficiently large, $P_e$ is close to one which means perfect anonymity. 

From \ref{eq:fanomi}, we see that mutual information plays a key role. It is given by the expectation
\begin{equation}
   \label{eq:mi}
    I(X;Y) = E \left [\log \frac{p(X\vert Y)\cdot p(Y)}{p(X)\cdot p(Y)}\right ] .
\end{equation}
It is zero if and only if  $p(X\vert Y)$ equals $p(X)$ for all $X=x$ and all $Y=y$. In other words: If the recipient $X$ is statistically independent of the observed data $Y$, (\ref{eq:mi}) becomes zero. In all other cases, we have a dependency, an information leakage that is captured with the conditional probability $p(X\vert Y)$. 

For example, suppose $Y$ represents the tunnel length, with a default length of $3$ units applied consistently accross the network. In this case, $p(y=3)=1$ and $0$ otherwise. Thus, it remains to determine the conditional probability distribution $p(X\vert y=3)$. 

Further observations such as tunnel lifetime, number of network participants, number of inbound and outbound tunnels per node, etc., complicate $p(X\vert Y)$ making it hard to model analytically. Even parametric models may lack enough expressiveness, requiring data-driven methods like \gls{ml} to  approximate the distribution. To this end we perform in Section \ref{experiments} computer experiments with synthetic and genuine I2P data that we feed into \gls{ml} models with the aim to "learn" from the data the conditional probability distribution $p(X\vert Y)$.

\subsection{Trustless and Anonymity}

In the context of communication protocols, trustless does not imply that the system is untrustworthy. Rather, it means that the system is designed to function correctly and securely without requiring trust in any individual participant, third party, or intermediary. Each node in the network may behave benignly or maliciously—and may even switch between these states at any time—yet the protocol remains secure by design, regardless of such behavior.

The core idea of "trustlessness" is to eliminate the middleman. It is rooted in cryptography, consensus mechanisms, transparency, and economic incentives. Cryptography uses digital signatures and hashing to prove ownership and ensure data integrity. Consensus protocols like Proof-of-Work and Proof-of-Stake enable strangers to agree on a single truth without a central authority, making cheating costly and honest behavior profitable. Transparency ensures all transactions and rules are public, allowing anyone to verify compliance through cryptographic proofs. Economic incentives align participants’ self-interest with network security by rewarding honesty and punishing malicious actions.

Key traits of a Trustless Protocol are:
\begin{itemize}
    \item No Privileged Parties: Rules and transactions cannot be unilaterally altered or reversed.
    \item Permissionless: Open to all without approval.
    \item Censorship-Resistant: Resistant to blocking valid transactions.
    \item Deterministic: Identical inputs yield consistent outputs.
\end{itemize}

I2P embodies trustless design by assuming an adversarial network with hostile nodes. It employs a form of onion routing bundling multiple encrypted messages so each node only knows its adjacent hops. Separate, unidirectional tunnels handle outbound and inbound messages, preventing links between, as an example, client requests and server responses. Its peer-to-peer topology eliminates central authorities, distributing tunnel management and maintaining implicit distrust of nodes. This architecture makes deanonymization computationally infeasible, as no single node can compromise user privacy.

Trustless is the foundation. It removes the need to rely on other participants or a central authority. Instead, trust is placed in code, cryptographic mechanisms, or open protocols. In a trustless network, one can operate anonymously — but one does not have to. However, if one wants to maintain anonymity the network must be trustless. This is because, within non-trustless (such as centralized or permissioned) networks, users must still rely on a trusted entity to safeguard their data, uphold privacy, and prevent the disclosure of their identity. 

We consider a passive attack scenario, where network traffic is recorded and analyzed using machine learning techniques to identify relevant patterns within large data volumes. Since the network is assumed to be trustless—meaning no node must be inherently trusted—useful patterns must be extracted directly from the collected data.

\section{Experimental Framework}
\begin{figure}[H] 
    \centering 
    \includegraphics[width=5in]{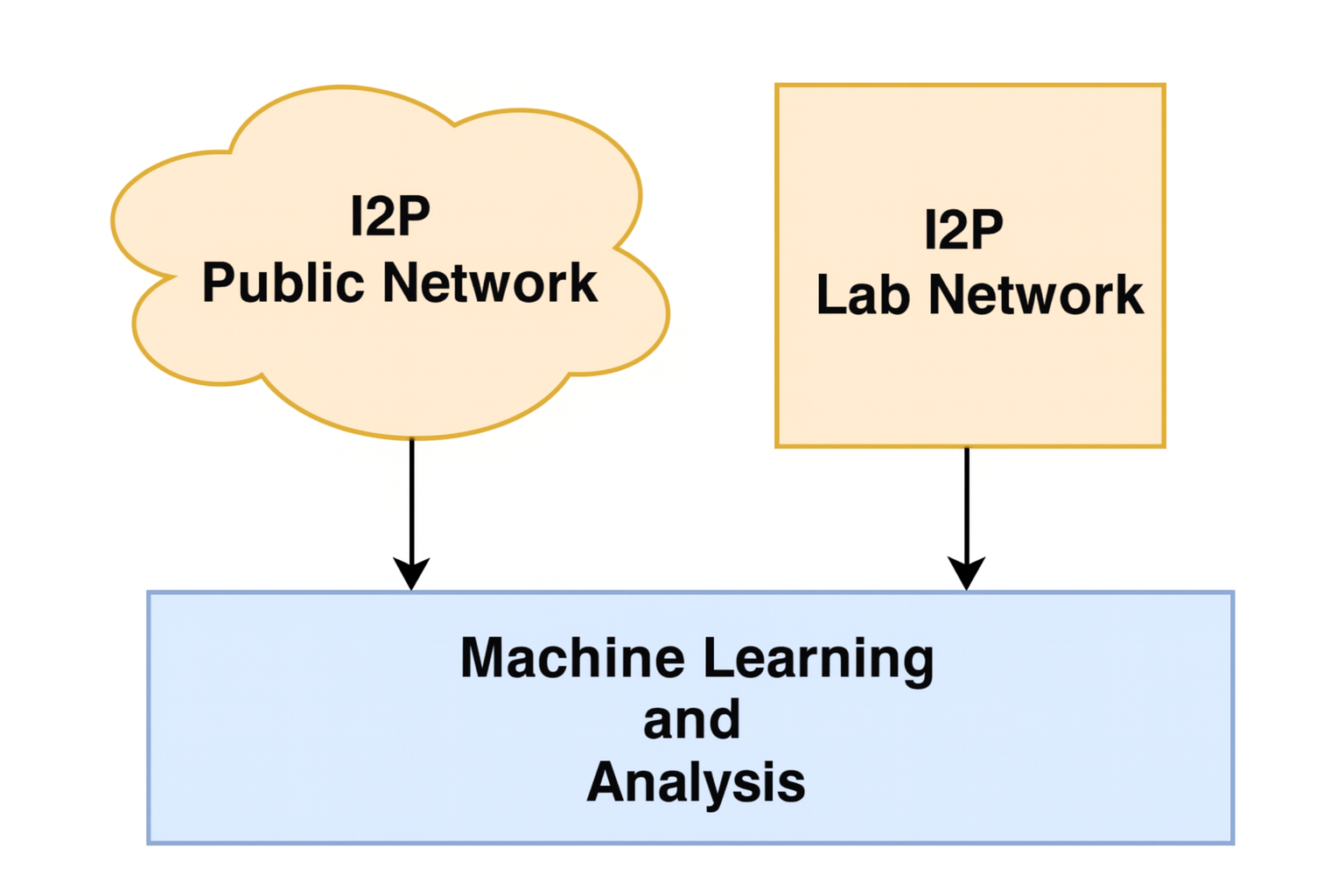} 
    \caption{Experimental framework.} 
    \label{fig:example} 
\end{figure}

We use an experimental framework in which all network traffic is recorded using tools such as \comp{tcpdump}. The recorded packets are automatically analyzed and prepared for ML models. Python is used as the programming language, supported by libraries such as TensorFlow for interactive analyses. The framework allows controlled experiments with HTTP requests and responses between \gls{i2p} nodes in the \gls{i2p} laboratory network (Fig. \ref{fig:example} right hand side) as well as analysis and/or validation of traffic from the real, public \gls{i2p} network (Fig. \ref{fig:example} left hand side).

\subsection{Laboratory Setting} \label{analysisnettraffic}
\begin{figure*}[h] 
    \centering 
    \includegraphics[width=\textwidth]{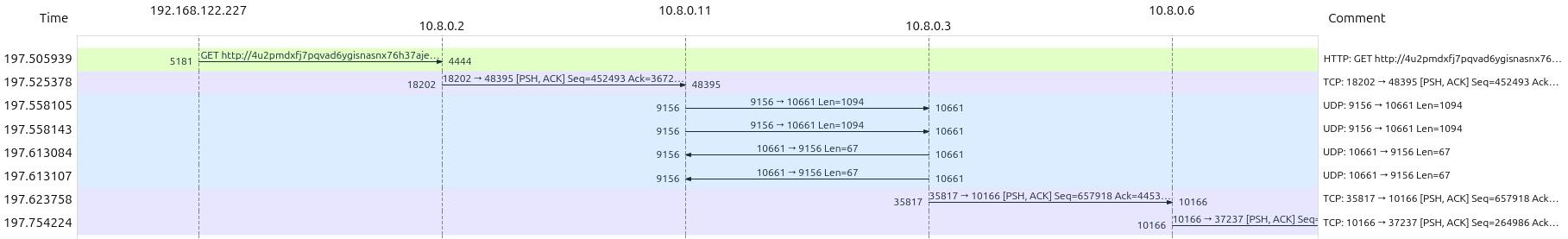} 
    \caption{Timing diagram from the sender 10.8.0.2 to the first node of the receiver's inbound tunnel 10.8.0.11.} 
    \label{fig:e2ei2p} 
\end{figure*}

The laboratory environment is isolated from the real, public network and used for generating synthetic network traffic. Multiple I2P nodes, each within its own Docker container, run on a single host. Figure \ref{fig:e2ei2p} shows the timing behavior and sequence of messages of a HTTP request from sender 10.8.0.2 to a receiver's input tunnel 10.8.0.11 in that setting. We clearly see that I2P transmits the packets over TCP and UDP which is an obfuscation strategy.

\begin{figure}[h] 
    \centering 
    \includegraphics[width=0.8\textwidth]{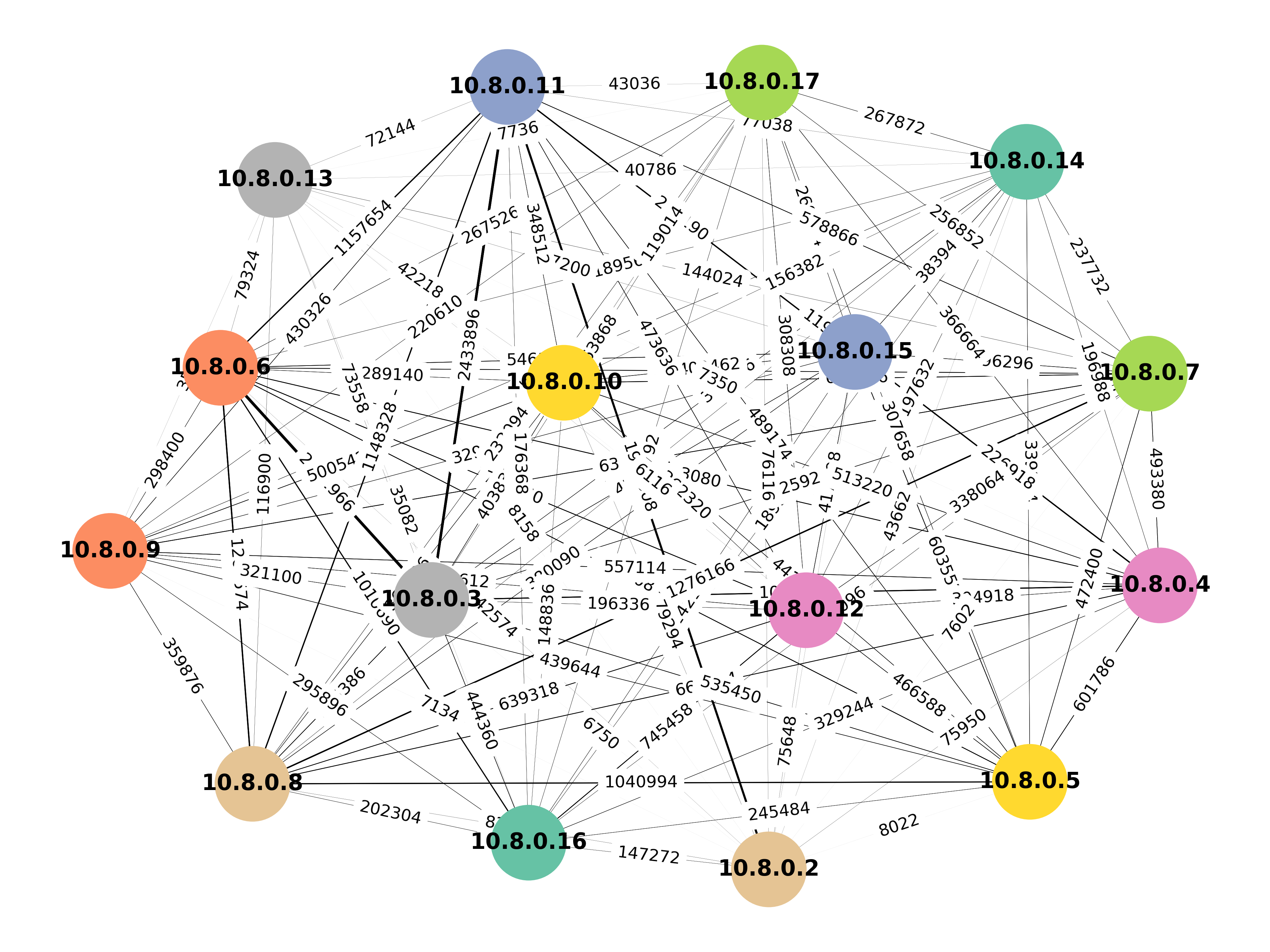} 
    \caption{Visualization of I2P network traffic based on all recorded Wireshark data.} 
    \label{fig:network} 
\end{figure}

Figure \ref{fig:network} visualizes the recorded data (several HTTP server requests, always from the same sender to the same recipient) and illustrates the I2P network with its 16 nodes in the laboratory network. Each node is color-coded (although multiple nodes may share the same color) and labeled with its corresponding IP address. The lines between the nodes represent the data traffic that occurred between them; the thicker the line, the more traffic was transmitted. Even in this minimized traffic scenario a highly meshed structure is already evident. Additionally, it's clear that a substantial amount of data traffic took place between nodes 10.8.0.2 and 10.8.0.11, as well as between 10.8.0.11 and 10.8.0.3.

\begin{figure}[h] 
    \centering 
    \includegraphics[width=0.8\textwidth]{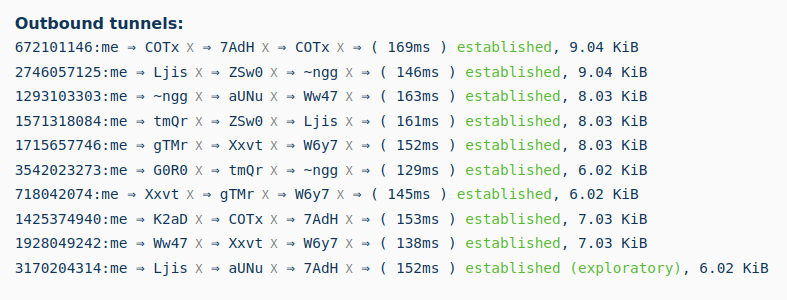} 
    \caption{Outgoing tunnels from the sender.}
    \label{fig:outbound} 
\end{figure}

Figure \ref{fig:outbound} shows the ten outbound tunnels of the sender. Each tunnel consists of the three hops. The last tunnel shown in that figure has status "established (exploratory)". This signifies that the tunnel has been successfully built. It is now active for use and the "exploratory" tag indicates that the tunnel's purpose is for network maintenance.


\begin{figure*}[t] 
    \centering 
    \includegraphics[width=\textwidth]{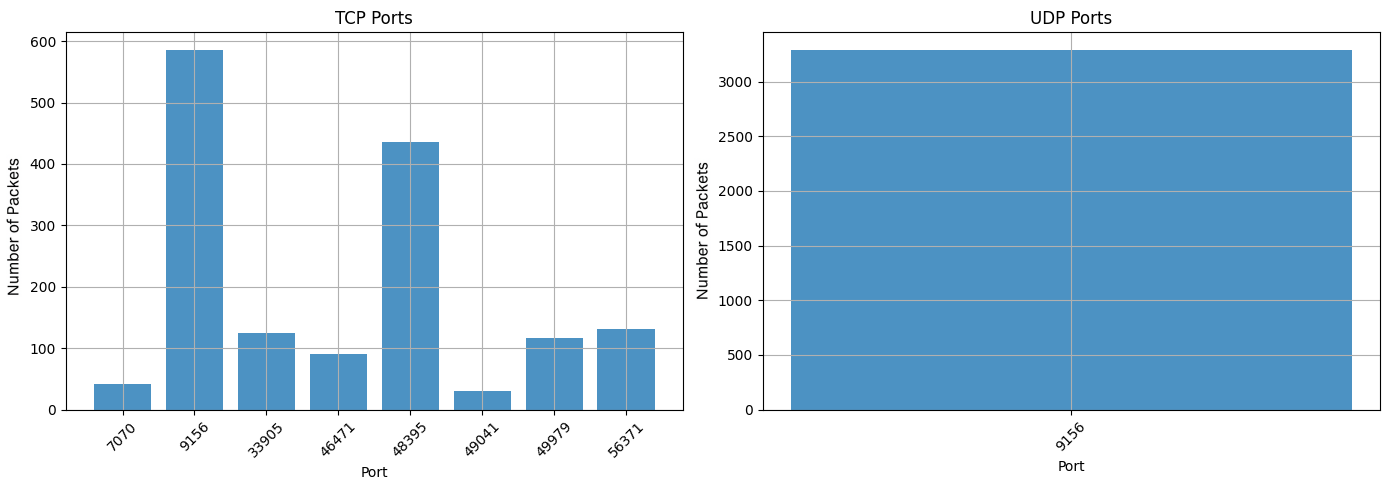} 
    \caption{Number of packets per port for TCP and UDP to node 10.8.0.11.}
    \label{fig:ports} 
\end{figure*}

The data from the tcp-dump were curated. All re-transmitted TCP packets ($\comp{retransmission}$ or $\comp{fast-retransmission}$) were excluded, as they had already been transmitted previously. Duplicate acknowledgments ($\comp{duplicate-ack}$), which do not carry new information, were also excluded. Additionally, all TCP packets without payload (length 0) were filtered out, leaving only relevant packets. Without this filtering, statistical evaluations could be distorted. Applying the filter reduced the number of recorded packets from 102,509 to 43,443.

To better understand how node 10.8.0.11 receives data, a breakdown per port for both UDP and TCP was created (see Fig. \ref{fig:ports}). The port number is always from the perspective of node 10.8.0.11 (first tunnel participant) for incoming communication. Most packets (3,286) arrive via UDP port 9156. The remaining 1,553 packets come via TCP. It is notable that port 9156 is used for both TCP (585 packets) and UDP (3,286 packets), making it the port with the highest activity overall. TCP port 48395 also shows high TCP activity with 435 packets; it is not used for UDP though. It becomes evident that all packets originate from TCP port 48395. The further analysis focuses now on that port with highest traffic.

\begin{figure*}[h] 
    \centering 
    \includegraphics[width=\textwidth]{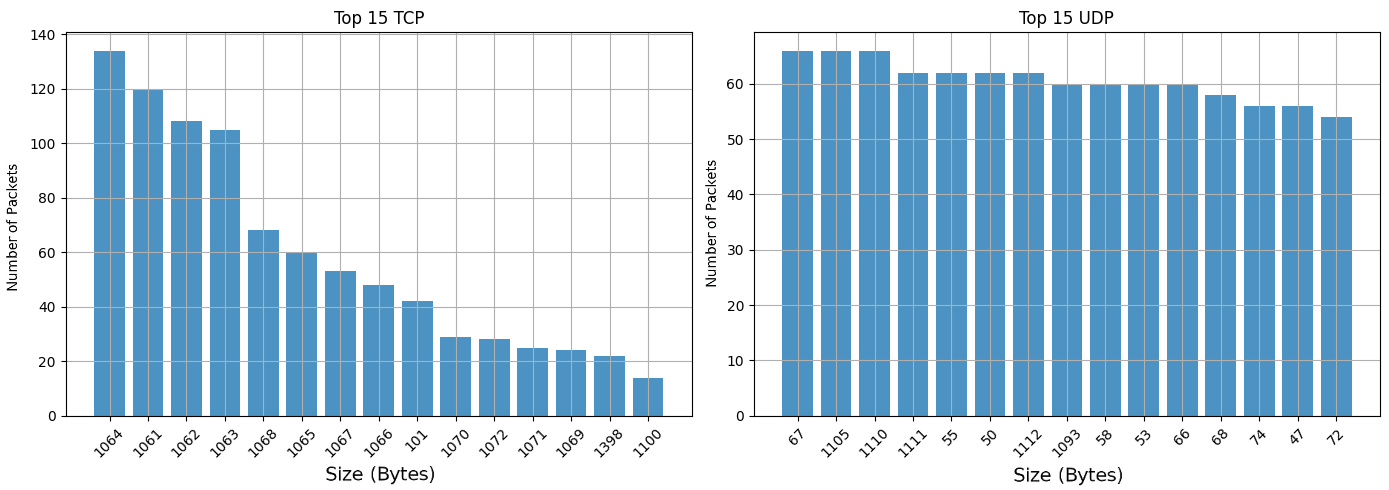} 
    \caption{Top 15 TCP and UDP packet sizes to node 10.8.0.11 }
    \label{fig:tcpudp} 
\end{figure*}

Next, the packet sizes (only the payload) arriving at node 10.8.0.11 were examined (see Fig. \ref{fig:tcpudp}). For TCP, it is evident that most packets are between 1,072 and 1,100 bytes in size\footnote{see "message preprocessing" in \url{https://geti2p.net/en/docs/tunnels/implementation}}. Outliers exist at 101 bytes and 1,398 bytes.
For UDP, the variation in packet sizes is more dispersed but consistent in packet count. Packet sizes vary greatly, from 47 bytes to 1,112 bytes. It is also evident that no packet exceeds the maximum size of 1,280 bytes (given by the routing mechanism in I2P). Smaller packets could be status packets, while the larger ones likely carry actual payload data.

The analysis from the first tunnel participant (10.8.0.11) shows that the payloads of the packets have a very high entropy close to 8 bits, indicating strong encryption and randomness (see Fig. \ref{fig:entropy} below).  The red dotted line represents the average entropy across all packets. The contents of the payload is thus not usable. The analysis bases mainly on metadata such as packet size, ports, and protocols. 
\begin{figure}[h!] 
    \centering 
    \includegraphics[width=0.8\textwidth]{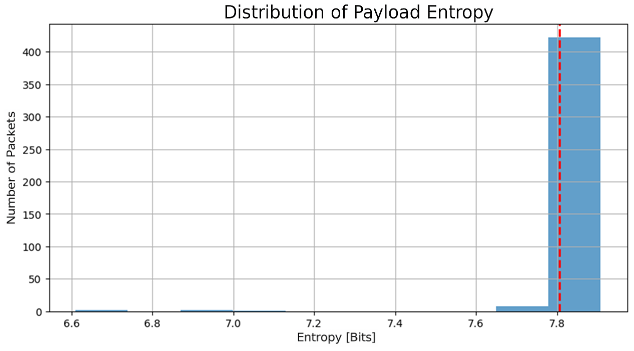} 
    \caption{Entropy of the payload of selected packets.}
    \label{fig:entropy} 
\end{figure}

\subsection{Real-world, public Network}
\begin{figure*}[h] 
    \centering 
    \includegraphics[width=\textwidth]{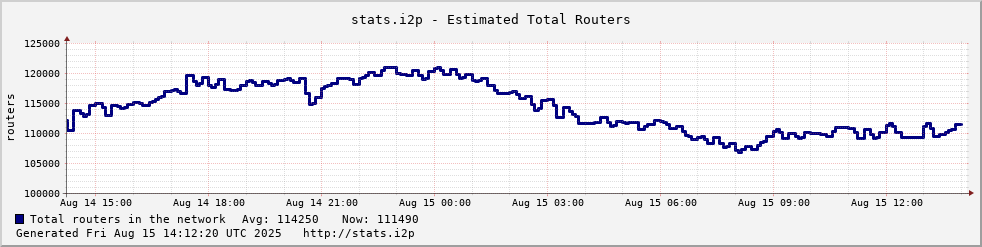} 
    \caption{Estimated number of I2P nodes over time worldwide.}
    \label{fig:routersWorldwide}
\end{figure*}

Now, we turn to the real-world, public I2P network. Figure \ref{fig:routersWorldwide} shows the estimated number of nodes in I2P worldwide over a 24 hours period and gives an indication of the network size. Clearly visible is the fluctuating number of network participants.

\begin{figure}[h] 
    \centering 
    \includegraphics[width=0.8\textwidth]{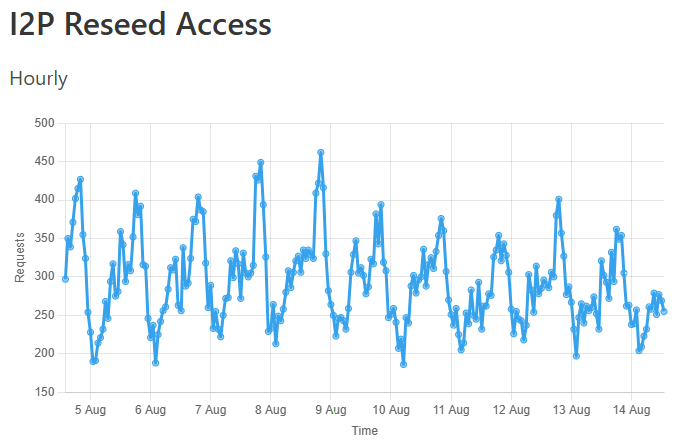} 
    \caption{Pattern of (re-)connecting I2P nodes.}
    \label{fig:patternReseed} 
\end{figure}

Figure \ref{fig:patternReseed} shows how one of the few reseed servers in the I2P network receives requests from I2P nodes over time. A clear 24-hour fluctuation pattern is visible, likely caused by I2P nodes in the U.S. time zone regularly leaving and rejoining the network.

In \cite{inria} and \cite{simioni} the authors investigated availability fingerprinting. By tracking up-time and network behaviors via NetDB and public endpoints, they could associate hidden services with specific routers. These works are related to the churn rate that measures the fraction of participants leaving/joining per time unit, regardless of whether they are the same or different peers. Adversaries struggle to track users across time and so a high churn rate usually protects against time-based attacks. However, if the same subset of participants churns, adversaries can ignore the churning group and focus only on the stable part of the network, a substantially reduced anonymity set.





\section{Experiments with ML methods } \label{experiments}

\gls{ml} is a key technology of artificial intelligence that enables the creation of models from training data that can then further be applied to unknown data. Supervised learning as subcategory of ML uses labeled data, while unsupervised learning recognizes patterns without labels. Deep Learning (DL) is a subcategory of artificial neural networks with multiple layers and automatically extracts features from raw data. Within that DL category, \gls{cnn} are particularly suitable for automatic feature extraction in complex data such as network traffic \cite{sakshi}, \cite{umargono}. The quality of the features is crucial for the success of the ML models.

Zhao et al. \cite{zhao} present state-of-the-art traffic classification methods that could be applied to identify \gls{i2p} traffic. The work summarizes advanced traffic analysis techniques, including statistical and behavioral patterns. The authors show how encrypted and protocol-obfuscated traffic can still be classified based on metadata and flow characteristics. Various features such as connection duration, destination port, or packet size were used. The payload-based methods achieved the best accuracy (often over 98$\%$), but lose effectiveness under encryption. 


The study showed also that none of the methods for feature extraction met all the requirements and it is highly dependent on a specific scenario. 
Since \gls{i2p} encrypts and obfuscates all traffic, manual feature engineering is very time-consuming. Therefore, a method with automatic feature extraction is needed. \gls{cnn} were chosen because they can automatically extract meaningful features from individual packets. Other models, such as \gls{rnn}, were not considered, since they rely on temporal sequences of packets. In \gls{i2p} traffic, however, each packet must be recognised individually, even if only a few packets from a request reach the node.


The following sub-sections present an analysis of four interconnected experiments investigating anonymity within the I2P network. Full de-anonymization was deemed practically impossible due to payload encryption. Instead, the focus was on classifying network data to link packets to specific services, a key step in forensic analysis. The experiments progressed from simple clustering of traffic patterns to advanced deep learning models detecting subtle correlations. The final validation tested these models in real-world conditions within the public I2P network, evaluating their forensic utility.

\subsection*{First Experiment: $k$-Means Clustering} 

This first experiment tested whether I2P network traffic shows basic, separable patterns. Using unsupervised $k$-Means clustering algorithm on features like port number, payload length, and protocol type,  results showed no natural groupings. The features marked with “Yes” in Table~\ref{tab:feature4kMeans} were selected for clustering. The dataset included all available packets, with the sender IP set to 10.8.0.2 and the receiver IP to 10.8.0.11, totaling 435 packets.

\begin{table*}[t]
\centering
\label{tab:wide}
\caption{Featuers fed into the $k$-Means Algorithm}
\begin{tabular}{l|c|c|c|c}
\textbf{Feature} & \textbf{Description} & \textbf{Type} & \textbf{Suitable} &\textbf{Comment}\\
\hline
IP-address  & IP address of sender   & categorical   & no  & Sender remains unchanged\\
Time  & Time when data packet is received   & numerical    & no  & Not enough information\\
Port  & Port of sender / recipient    & categorical     & yes  & -\\
Length of payload  & Length of payload   & numerical    & yes  & Similar length, similar contents\\
Payload & Payload itself   & numerical    & yes  & - \\ 
Protocol  & TCP or UDP   & categorical    & yes  & I2P uses UDP and TCP\\
\end{tabular}
\label{tab:feature4kMeans}
\end{table*}

The elbow method is a common technique for selecting the optimal number of clusters in $k$-Means clustering. It involves plotting the distortion (measured as within cluster sum of squares) against $k$ and identifying the point where further increases in $k$ result in only minor reductions in distortion. This "elbow point" represents a good balance between model complexity and clustering performance.

\begin{figure}[h] 
    \centering 
    \includegraphics[width=0.8\textwidth]{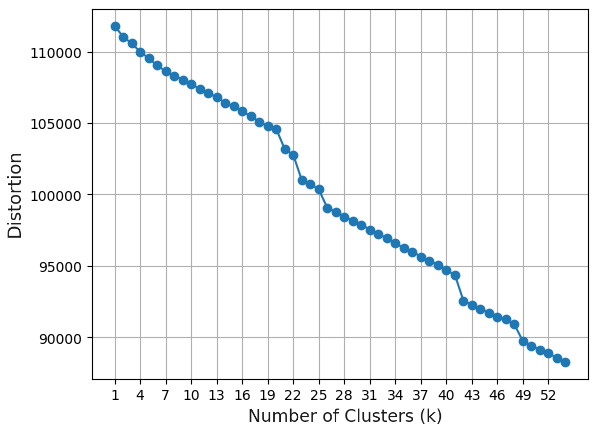} 
    \caption{Elbow Method for Determining the Optimal Number of Clusters}
    \label{fig:elbow} 
\end{figure}

In Figure \ref{fig:elbow}, the distortion - measured as root mean square squared - decreases almost linearly with increasing $k$, showing no clear elbow. This suggests that the data does not have a natural number of clusters, and there is no clear rule for selecting an optimal $k$ in this case.

Let us focus on $k = 20$ clusters, as a slight change is visible in Figure~\ref{fig:elbow} at this point. Among the 20 clusters, one cluster contained 75 data points, up to four clusters had only 3 data points each, and the average cluster size (excluding the largest) was 22.89 data points.


The experiment confirmed that simple unsupervised methods such as $k$-Means are insufficient to derive meaningful insights into communication within the I2P network. The data structure is too complex and noisy to be divided into informative clusters using basic distance metrics. The failure of this fundamental approach underscored the necessity of more advanced, supervised learning methods capable of detecting hidden and complex patterns—leading directly to the conception of the second experiment.

\subsection*{Second Experiment: CNN as Classifier}

Following the failure of unsupervised clustering, the approach shifted to a supervised deep learning model—a Convolutional Neural Network (CNN)—to exploit its capability for automatic feature extraction and binary classification of traffic patterns. The central research question was: “Can a CNN reliably classify I2P traffic into two categories—relevant HTTP packets and irrelevant background traffic?”

Building on this, the experiment aimed to evaluate whether the CNN could also differentiate between traffic directed to different recipient nodes. In this extended setup, Class 1 represents irrelevant traffic unrelated to the request, Class 2 corresponds to packets intended for the first tunnel node, and Class 3 denotes packets directed to an alternative tunnel node.

\begin{figure}[H] 
    \centering 
    \includegraphics[width=0.35\textwidth]{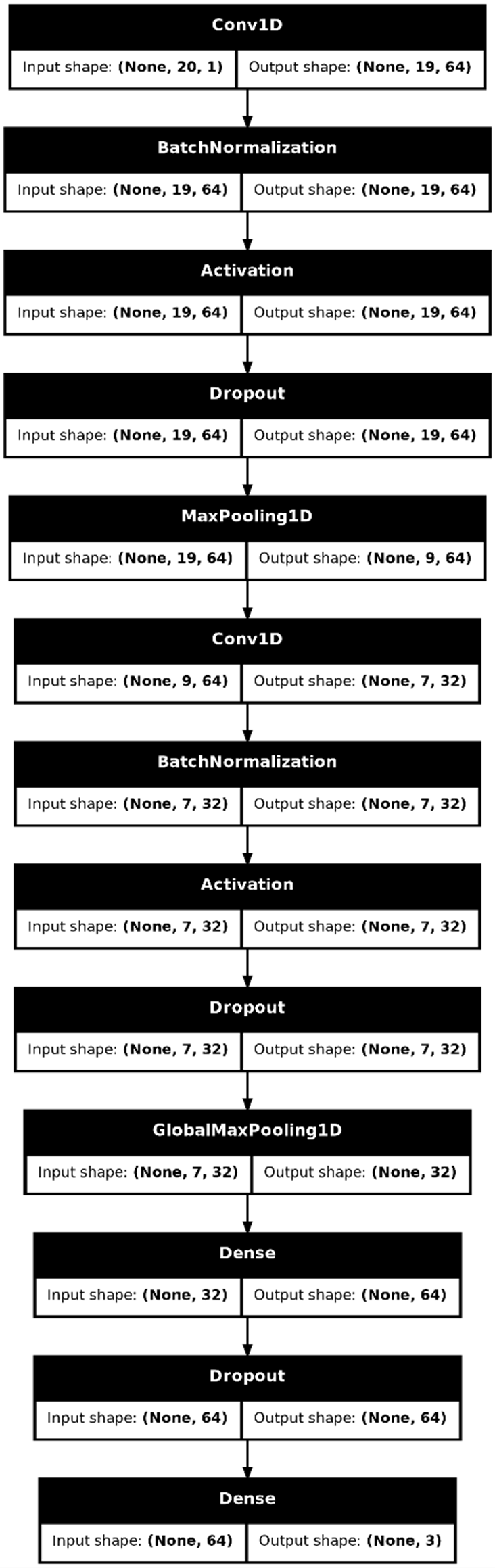} 
    \caption{Architecture of the CNN}
    \label{fig:cnn} 
\end{figure}

The implemented CNN architecture consisted of two convolutional layers, each incorporating batch normalization and ReLU activation. Dimensionality reduction was achieved through pooling operations, followed by a global pooling layer to summarize feature maps. The network concluded with a dense layer and a sigmoid-activated output for binary classification (see Figure \ref{fig:cnn}, Appendix). Source code is available at \url{https://anonymous.4open.science/r/i2pnodesniffer-5BBC/README.md} .


The CNN was trained to classify network packets into HTTP requests and other types of traffic. Two datasets were used for the experiments: a smaller one containing 101 requests and a larger one with 1,000 requests (see Table~\ref{tab:second_experiment_dataset}). Only requests that could be unambiguously verified by an automated validation script were included. This restriction was necessary to ensure that each packet could be reliably associated with a specific request. Consequently, the number of samples “Used for Training” is not exactly 200 and 2,000, respectively.

\begin{table}[H]
\centering
\caption{Datasets used for the experiment}
\begin{tabular}{c|c|c|c}
\textbf{Dataset} & \textbf{Total Packets} & \textbf{Used for Training} & \textbf{Used for Validation} \\
\hline
Small  & 3432   & 166   & 20  \\
Large  & 58,088 & 1598  & 200 \\
\end{tabular}
\label{tab:second_experiment_dataset}
\end{table}

The correlation between each input features and the predicted value was analyzed. The analysis of this correlation is shown in Table \ref{tab:correlation}. It demonstrates that the individual input features only have a very weak correlation with the target variable / target class, indicating that they have little explanatory power on their own. This is in agreement with the results from the $k$-Means algorithm above. 

\begin{table}[H]
    \centering
    \caption{Correlation between the Prediction and the Input Feature}
    \begin{tabular}{c|c}
        \textbf{Feature}	    & \textbf{Correlation}  \\
         \hline
         $\mathrm{tcp\_seq}$ &	0.102051  \\
         $\mathrm{dst\_port}$ & 0.082602 \\
         $\mathrm{payload\_size}$ & 	0.020137 \\
         $\mathrm{ip\_len}$	& 0.020137 \\
         $\mathrm{ip\_ttl}$	& 0.018280 \\
         $\mathrm{tcp\_dataofs}$ &	-0.000901 \\
         $\mathrm{old\_frame\_num}$	&-0.002678 \\
         $\mathrm{frame\_num}$	&-0.002693 \\
        $\mathrm{timestamp}$	& -0.002700 \\
         $\mathrm{tcp\_window}$	& -0.033774 \\
         $\mathrm{tcp\_ack}$	& -0.037385 \\
         $\mathrm{tcp\_flags}$	& -0.043201 \\
         $\mathrm{src\_port}$	& -0.101007 \\
         $\mathrm{ip\_id}$	& -0.138102 \\
    \end{tabular}
    \label{tab:correlation}
\end{table}
During CNN training, an equal number of elements were selected from each class to prevent any single class from dominating the learning process.
 
The accuracy reported in the experiments refers to a subset of packets extracted from the dataset prior to training. This subset was used to assess how well the model classifies previously unseen packets. A separate portion of the recorded data, which was not used during training, served as the evaluation set. Model performance was measured by accuracy, defined as the proportion of correctly classified packets. The experiment confirmed the hypothesis that no meaningful conclusions about the communication can be drawn.

The results are shown in Table~\ref{tab:second_experiment_result_big_dataset}. The CNN achieved high accuracy on the validation data set (up to 99.5$\%$), particularly when metadata were used. In comparison, the variants using "Payload only" as well as the variant "All raw data" exhibited poor performance and high variance in accuracy. The classification of all packets showed a significant deterioration, indicating overfitting to training data. These results also underscore the strength of CNNs, which are capable of uncovering complex, nonlinear, and hidden patterns in data.
\begin{table}[H]
\centering
\caption{CNN test variants with large data set containing 200 packets for manual validation}
\begin{tabular}{l|c|c}
\textbf{Variant} & \textbf{Correctly Classified} & \textbf{Accuracy} \\ \hline
Payload only      & 145 -- 153 & 72.50\% -- 76.50\% \\ 
All raw data      & 181 -- 184 & 90.50\% -- 92.50\% \\ 
Without payload   & 199 & 99.5\% \\ 
Without port      & 183 -- 187 & 92.50\% -- 93.50\% \\ 
\end{tabular}
\label{tab:second_experiment_result_big_dataset}
\end{table}
Remarkably, the “Without Payload” variant achieved nearly perfect classification accuracy (99.5 $\%$) on the large dataset. This strongly suggests that the decisive classification patterns lie in the metadata rather than the encrypted content.

Despite impressive validation accuracy, applying the models to the entire recorded dataset revealed a significant performance drop. While the “All Raw Data” model performed poorly (64.68$\%$ accuracy for Class 1), the “Without Payload” model maintained high accuracy (95.17$\%$). This indicates that encrypted payloads introduced noise that confused the model, whereas metadata provided a clean, learnable signal under controlled conditions. The models particularly struggled to identify the majority class of irrelevant traffic (Class 1), leading to an unacceptable rate of false-positives as shown in Table ~\ref{tab:second_experiment_result_all_dataset} .


\begin{table}[H]
    \centering
    \caption{Classification results of all recorded data (large dataset) using different feature variants}
    \begin{tabular}{l|c|c|c}
        \textbf{Variant} & \textbf{All data} & \textbf{Without payload} & \textbf{Without port} \\  \hline
        Class 1 & 37,009 & 54,424 & 44,848 \\ 
        Class 2 & 21,079 & 3,664 &  13,240\\ \hline
        Accuracy of Class 1 & 64.68$\%$  & 95.17$\%$ & 78.27$\%$    \\ 
        Accuracy of Class 2 & 97.78$\%$     & 100$\%$  & 90.32$\%$ \\ 
    \end{tabular}
    
    \label{tab:second_experiment_result_all_dataset}
\end{table}

In order to further investigate the relevance of individual features, various combinations of raw data columns were evaluated using the CNN. The results showed that not all features or combinations thereof had the same impact on classification performance. For example, a representative combination included the following columns (see Table~\ref{tab:accuracy}) \comp{src\_port}, \comp{dst\_port}, \comp{payload\_size}, \comp{tcp\_ack} and \comp{tcp\_seq}. This combination was analyzed to determine how removing individual columns would affect accuracy. It was found that classification accuracy could decrease significantly depending on the combination. When both \comp{tcp\_ack} and \comp{tcp\_seq} were omitted at the same time, accuracy dropped to only 69.33\%.

\begin{table}[H]
    \centering
     \caption{CNN Test Variants using Different Feature Combinations}
    \begin{tabular}{c|c|c}
        	\textbf{With/without pay-load} &	\textbf{Missing rows}	&\textbf{Accuracy} \\ \hline
         without	& $\mathrm{tcp\_ack}$, $\mathrm{tcp\_seq}$	&69.33$\%$ \\  
        without	& $\mathrm{dst\_port}$	&70$\%$ \\  
        without	& $\mathrm{payload\_size}$ &	69$\%$ \\  
         with	& none	& 63.67$\%$ \\  
    \end{tabular}
    \label{tab:accuracy}
\end{table}

In summary, CNNs can achieve unexpectedly high classification accuracy under controlled lab conditions, with metadata proving especially effective. However, the critical weakness lies in the lack of generalization to unseen, larger datasets. This raised the next logical question: If binary classification works in the lab, can the model also distinguish between multiple services or destinations?

\subsection*{Third Experiment: CNN for Multi-Class Classification }

Building upon the promising yet ambiguous binary classification results in the second experiment, the third experiment increased complexity. It aimed to test whether the CNN could not only separate relevant from irrelevant traffic but also differentiate between two distinct HTTP destinations.

The central question was: “Can the CNN model correctly assign packets to one of three classes—irrelevant traffic (Class 1), HTTP request to Target A (Class 2), and HTTP request to Target B (Class 3)?” The methodology followed that of Experiment 2 but with three classes instead of two.
The third experiment tests whether the CNN can distinguish between two different recipient nodes.

Manual validation on a separate test dataset again yielded superficially high accuracy values, particularly for variants including metadata, see results in Table \ref{tab:second_experiment_result_big_dataset_with_all_recorded_data}. 

As in Experiment 2, performance collapsed when classifying all recorded packets. Again, the “Without Payload” model outperformed the others: it classified the dominant irrelevant traffic class (Class 1) with 81.14$\%$ accuracy, while the “All Data” model achieved only 57.11$\%$. Although validation results appeared promising, the model struggled to reliably distinguish services once confronted with the much larger, unfamiliar dataset.

In contrast, the other variants showed significant misclassifications, suggesting that combining payload information increases the complexity of the classification task for the model.
\begin{table}[H]
\centering
\caption{Recorded Data Classified with the Trained Models. Remark: Classes~1--3 indicate the number of packets assigned to the respective class.}

\begin{tabular}{l|c|c|c}
        \textbf{Variant} & \textbf{All data} & \textbf{Without payload} & \textbf{Without port} \\ \hline
        Class 1 & 56,707 & 80,199 & 80,672 \\ 
        Class 2 & 37,681 & 9,852 &  9,706\\ 
        Class 3 & 6,258 & 10,595 &  10,268\\ \hline
        Accuracy Class 1 & 57.11\%  & 81.14\% & 81.33\%    \\ 
        Accuracy Class 2 & 57.11\%  & 100\%  & 66.74\% \\ 
        Accuracy Class 3 & 73.01\%  & 99.89\%  & 93.14\% \\ 
\end{tabular}

\label{tab:second_experiment_result_big_dataset_with_all_recorded_data}
\end{table}

The findings indicate that while the model performed well under lab conditions, it was not robust enough in practice to distinguish service classes reliably. The patterns learned during training were too specific to the controlled environment and did not generalize to the more heterogeneous traffic. This led to the final and decisive test—transferring the lab-trained models to the uncontrolled public I2P network.

\subsection*{Fourth Experiment: Generalisation to Public I2P Network}

This final experiment was crucial to assess if high-performing lab models had genuine forensic applicability or if their success was limited to the artificial lab setting.

The models trained in the lab were applied to data from the public I2P network. To capture real-world traffic, an I2P service was deployed via Docker on a host system, with requests sent from a separate client. Network traffic was recorded using tcpdump on both the host and within the container.

Table \ref{tab:fourth_experiment_result} shows results with many misclassifications; the results are mostly unusable. Instead of 15 Class 2 packages, 71.6 to 88.4 times more "Without port" packets were incorrectly labeled as such. The variant "Without payload" fared slightly better.

\begin{table}[H]
\centering
\caption{Recorded Data Classified with the Trained Models (Host System vs. Docker Environment).}
\begin{tabular}{|c|l|c|c|c|c|}
\hline
\textbf{No.} & \textbf{Variant} & \textbf{Class} & \textbf{Host} & \textbf{Docker} & \textbf{Multiple} \\ \hline
\multirow{2}{*}{1} & \multirow{2}{*}{Without port} 
& 1 & 2103 & 1807 & - \\ \cline{3-6}
 &  & 2 & 1228 & 1326 & 81.8 -- 88.4 \\ \hline
\multirow{2}{*}{2} & \multirow{2}{*}{Without payload} 
& 1 & 3138 & 2935 & - \\ \cline{3-6}
 &  & 2 & 193 & 198 & 12.8 and 13.2 \\ \hline
\multirow{2}{*}{3} & \multirow{2}{*}{Payload only} 
& 1 & 2233 & 2058 & - \\ \cline{3-6}
 &  & 2 & 1098 & 1075 & 71.6 and 73.2 \\ \hline
\end{tabular}
\label{tab:fourth_experiment_result}
\end{table}

The conclusion is that the lab-trained models failed in the public I2P network. This demonstrates that patterns learned in a homogeneous lab environment were not generalizable to the complex and diverse real-world I2P traffic.

\subsection*{Discussions}
Synthesizing the findings from all four experiments provides a clear answer to the overarching research question on the reducibility of anonymity in the I2P network through passive traffic analysis. The gradual escalation in complexity—from simple clustering to deep learning and real-world testing—yielded a consistent picture.

The key insights can be summarized as follows:
\begin{itemize}

\item Failure of Simple Methods: Classical unsupervised clustering techniques such as k-Means are inadequate for identifying meaningful patterns in heavily obfuscated I2P traffic, which lacks simple, separable structures.

\item Potential of Deep Learning under Lab Conditions: CNNs can achieve high classification accuracy in controlled, homogeneous lab environments, learning patterns invisible to simpler algorithms.

\item Dominance of Metadata: The most significant insight is that packet metadata (e.g., ports, sizes, timing) provided far more effective signals for classification than encrypted payloads. In all lab experiments, models trained solely on metadata consistently and significantly outperformed those using payload data, which acted as confusing noise.

\item Lack of Generalizability: The decisive weakness—and final outcome—was the complete failure of these models when transferred to real-world conditions. Patterns learned in the lab did not generalize to the heterogeneous and dynamic nature of public I2P traffic.

\end{itemize}

In summary, the methods examined for reducing anonymity in the I2P network through passive traffic analysis are unsuitable for practical forensic use. The successes achieved under laboratory conditions could not be replicated in the real world, underscoring the effectiveness of I2P’s protection mechanisms against such analytical attacks.

\section{Conclusions}

This study examined the potential for deanonymizing services within the \gls{i2p} network through passive traffic analysis and machine learning. We began by outlining the fundamental principles of anonymity in mix networks and demonstrated how \gls{i2p} extends these principles through unidirectional tunnels that enhance anonymity and support robust communication in trustless environments. It makes clever usage of public and private keys on application layer to guarantee end-to-end confidentiality, while anonymity is achieved at the transport layer by limiting communication knowledge to immediate neighboring nodes.

Fano’s inequality was employed to theoretically characterize the primary sources of information leakage. Because a closed-form estimation proved intractable, a machine learning–based approach was adopted to approximate potential leakage and identify latent data-driven patterns. Experimental results confirmed that the payload in the \gls{i2p} network remains strongly encrypted and does not reveal direct information about the underlying services. Classical clustering algorithms such as $k$-Means proved ineffective, while Convolutional Neural Networks (CNNs) demonstrated only limited pattern-recognition capability in a controlled laboratory setting. Metadata emerged as the most informative feature set, whereas encrypted payloads contributed minimal analytical value. Attempts to generalize these models to the public \gls{i2p} network were unsuccessful, likely due to the high variability and dynamic nature of real-world traffic.

A complete deanonymization of services could not be achieved. Although anonymity networks like \gls{i2p} effectively resist direct exposure, they may remain vulnerable to persistent, large-scale statistical attacks. Possible adversarial strategies include controlling multiple nodes to filter or correlate packets, or deliberately disrupting network operations to isolate relevant communication paths.

Future research could explore hybrid architectures that combine Long Short-Term Memory (LSTM) networks, capable of modeling temporal aud causal dependencies at the application layer, with CNNs analyzing transport-layer features. Whether such models can provide deeper insights into \gls{i2p} traffic remains an open question.

In summary, this work provides (1) a clear conceptual description of \gls{i2p} and its anonymity mechanisms, (2) a theoretical framework for analyzing information leakage using Fano’s inequality, and (3) empirical evidence from both synthetic and real-world traffic demonstrating that deanonymization through CNN-based methods is not feasible under current conditions.

\section*{Acknowledgement}
\addcontentsline{toc}{section}{Acknowledgement}
We are very thankfull to Lance James from Unit 221B for a review of this preprint.  In particular, we are grateful for the exchange with Lance on David Chaum’s Dining Cryptographers Problem, the role of mix-nets as a foundational concept for I2P from its inception, and I2Ps’s independent origins from Tor.







\section*{Appendix A - CNN Architecture}
In \gls{cnn}, the \textit{input shape} specifies the structure of the data fed into the network. It is commonly represented as \((\text{batch size}, \text{sequence length},\text{number of features})\).

%
\newpage
\section*{Appendix B - CO2 Emission Related to Experiments}

Responsible use of LLMs: the study authors used ML models because it was evident from former studies that network traffic can be classified using ML models. Based on this knowledge, it is a reasonable assumption that ML models also might be useful to classify network traffic related to specific services within anonymous networks. 

All computer experiments were done on a private, local infrastructure as specified in Table \ref{tab:compres}. Trainings required approximately 120 hours. No LLMs were computed / used.

\begin{table}[htbp]
\centering
\caption{Computing Resources}
\label{tab:system_resources}
\begin{tabular}{ll}
\hline
\textbf{Host-System} & \\ 
\hline
OS & Ubuntu 24.04.2 LTS \\
CPU & AMD Ryzen 7 PRO 6850 with Radeon Graphics \\
CPU(s) & 16 \\
Memory & 19 GiB \\
\hline
\textbf{VM-System} & \\ 
\hline
OS & Ubuntu 22.04.2 LTS \\
CPU(s) & 6 \\
Memory & 8 GiB \\
\hline
\end{tabular}
 \label{tab:compres}
\end{table}

\end{document}